\documentclass[twocolumn,aps,prl,showpacs,amsmath,
amssymb,superscriptaddress]{revtex4}
\usepackage{graphicx}
\usepackage{dcolumn}
\usepackage{bm}

\begin{document}
\def\be{\begin{equation}}
\def\ee{\end{equation}}
\def\bearr{\begin{eqnarray}}
\def\eearr{\end{eqnarray}}
\def\tc{$T_c~$}

\title{Resonating Valence Bond Mechanism of\\
Impurity Band Superconductivity in Diamond}

\author{G. Baskaran}
\affiliation{Institute of Mathematical Sciences\\
Chennai 600 113, India}
\affiliation{IFCAM \& 
Institute for Materials Research\\
Tohoku University\\
Sendai 980-8577, Japan}

\begin{abstract}
Superconductivity in an uncompensated boron doped diamond, a very recent 
observation, is strikingly close to an earlier observation of Anderson-Mott 
insulator to metal transition, prompting us to suggest an electron correlation 
driven superconductivity in an impurity band. Random coulomb potential remove 
a three fold orbital degeneracy of boron acceptor states, resulting in an 
effective single, narrow, tight binding and half filled band of holes. 
Singlet coupling between spins of neighboring neutral acceptors $B^0-B^0$ 
is the seed of pairing. Across the insulator to metal transition, 
a small and equal fraction of charged $B^+$ and $B^-$ states (free
carriers) get spontaneously generated and delocalize. Thereupon neutral 
singlets resonate and get charged resulting in a resonating valence bond 
(RVB) superconducting state.
\end{abstract}

\maketitle
Diamond has been reported\cite{ekimovNature1} to become a superconductor 
upon high boron doping with a \tc $\approx 4~K$. This remarkable 
discovery of superconductivity in an uncompensated p-type semiconductor 
has possible implications for basic science and technology. As an example, 
shallow dopant states in Si and Ge play a key role in modern solid state 
electronics; from physics point of view they offer a testing ground for 
various ideas such as Anderson and Mott localization and their interplay.
Boron doped diamond\cite{ramdas}, being a relatively deep level acceptor 
with a hole binding energy of $\approx 0.37~eV$ provides a new energy scale, 
new possibilities\cite{may} and a rich physics. Further investigation of 
this system should be rewarding.

The present problem of electronic phases of boron doped diamond, 
containing competing coulomb interaction, randomness and coupling
to phonon, is like Si:P, quite a complex and challenging many body 
problem\cite{RMPreview}. However, using a body of insights one has 
gained in the last couple of decades from theory and experiments one 
can suggest physically plausible mechanism and insights, which could be 
guiding hypothesis to understand the observed superconductivity in boron 
doped diamond (diamond:B). It is in this spirit, the present letter suggests 
a minimal model and mechanism for superconductivity. The dopant density 
seems\cite{prins} to be just above the critical density of the 
Anderson-Mott transition making electron correlation important. We suggest 
a resonating valence bond (RVB) mechanism of 
superconductivity\cite{pwaScienceBZAgauge,pwaVanilla,gbrvbms}
in boron impurity band system. Diamond, a broad band insulator, where 
electron correlations are not important, should be viewed as offering an 
appropriate {\bf vacuum} to boron (atom with an odd electron number) subsystem, 
where electron correlation driven Anderson-Mott insulator to superconductor 
transition and other rich correlation physics could take place(figure 1).
\begin{figure}
\includegraphics*[width=8.0cm]{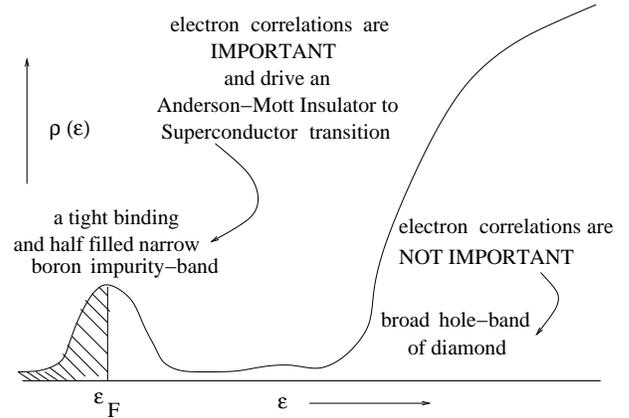}
\caption{\label{fig1} 
Hole density of states (schematic) in boron doped diamond, an uncompensated 
p-type semiconductor. Holes of acceptors form a strongly correlated and 
effective single impurity band at half filling. Anderson-Mott insulator to 
superconductor transition is suggested to take place in the impurity band 
as we increase boron density (figure 2).}
\end{figure}

The observed superconductivity \cite{ekimovNature1} in diamond:B
 is of type-II with a 
\tc $\approx 4~K$, $H_{c2} \geq 3.5~T$ and coherence 
length $\xi_{\rm GL} \sim  100~Au$. The dopant boron concentration is around 
$4.5 \times 10^{21}/cm^3$. 

Transport properties of diamond:B has been
studied by various authors and the issue of metal insulator transition
in the impurity band discussed\cite{prins, bustarret}. 
The only report to my knowledge, which makes a low temperature 
measurement of the critical boron concentration 
by a scaling analysis is by Tshepe et al.\cite{prins}. They find a critical 
doping concentration $n_c \approx 4 \times 10^{21}/cm^3$, strikingly close 
to the boron concentration $\sim 4.5 \times 10^{21}/cm^3$ used in
the new superconductor\cite{ekimovNature1}. It is this closeness which 
prompted us to examine the 
role of electron correlation and suggest a Anderson-Mott insulator to 
RVB superconductivity transition, similar to the one advocated 
recently\cite{gbrvbms, fczhang} in the context of (non-random) Mott 
insulator to superconductor transition.  Importance of strong correlation
is further enforced by a nearly temperature independent and a very 
large value of normal state resistivity of the (polycrystalline) 
superconducting sample\cite{ekimovNature1}, $\rho \sim 9~m\Omega~cm$, which 
gives a mean free path $\ell_0 \sim 0.1~Au$, much less than the average 
separation between neighboring acceptor sites. Resistivity in single 
crystal or low frequency conductivity measurements in polycrystalline 
sample will be welcome to substantiate this point.  

Diamond, an excellent insulator has a very broad band and a wide gap 
$\sim 5.6~eV$. The top of the valence band is three fold degenerate with 
effective 
masses, $2.12~m_e,~ 0.7~m_e~$ and $ 1.06~m_e $. Boron, which has 
one less electron compared to a carbon, becomes an acceptor, when it 
substitutes a carbon atom in diamond\cite{ramdas}. The acceptor states 
are threefold degenerate with a hole binding energy of 
$E_B \approx 0.37~eV$.  Spin-orbit coupling partly removes this degeneracy 
by a marginal $ 6~meV$. 

The relative dielectric constant of diamond is $\epsilon_0 \sim 5.6$. 
Effective mass theory estimate of acceptor states gives a hydrogenic 
impurity state with an effective Bohr radius of $a_B^* \equiv
\frac{e^2}{2\epsilon_0 E_B}\sim 3~Au$, of the `envelope function'. 
When this is used in conjunction with Mott's criterion of Mott insulator 
metal transition, $n_c a^3 \approx 0.25$, one obtains\cite{sato} 
a critical doping, $n_c \approx 8.0 \times 10^{21}/cm^3$. 
Experimentally observed critical concentration of reference [7] 
is of the same order as this rough estimate. 

Keeping the above in mind, we build a simple tight binding model for
the impurity band. The three fold degenerate acceptor wave functions 
(transforming as $t_{2g}$) have different spatial density distribution 
$|\psi_{\alpha}({\bf r} - {\bf R}_i)|^2$,for  $\alpha = 1,2$ and $3$.
Here ${\bf R}_i$ is the site of a boron atom. Hence the total 
electrostatic potential a hole 
in an acceptor state feels from neighboring dopants depends on the orbital 
it occupies. In other words, the three fold degeneracy of an acceptor 
state is in general lifted by the randomness. Simple estimate shows that
the amount of lifting, for the boron doping of interest, is large compared
to the impurity band width $ \sim 0.2~eV$. For low energy physics of
interested to us, only the lowest energy acceptor states are important. This 
is also consistent with general experience in Si:P, where the degeneracy 
of donor states has no substantial role and a non-degenerate donor state 
theory seem to work well\cite{bhat}.

Once we identify a lowest energy acceptor state for each boron, a one 
band tight binding model Hamiltonian follows:
\bearr
H & \approx & \sum_i \epsilon_i c^{\dagger}_{i\sigma} c^{}_{i\sigma}
-\sum t_{ij} c^{\dagger}_{i\sigma} c^{}_{j\sigma} + h.c. \nonumber \\
& + & \sum_i U_i (1-n_{i})^2
+ {1\over 2} \sum W_{ij} (1-n_{i})(1-n_{j})
\eearr
Here $c^{\dagger}_{i\sigma}$ is the hole creation operators at the lowest 
energy acceptor state of boron at site i with an energy $\epsilon_i$ 
and $t_{ij}$ are the hopping matrix elements; $n_i \equiv 
n_{i\uparrow} + n_{i\downarrow}$. The parameter $U_i$ is 
the Hubbard repulsion term of an acceptor state centered at site i 
and $W_{ij} \approx \frac{e^2}{\epsilon_0 R_{ij}}$ is the diagonal 
coulomb matrix element between two acceptor states separated by a 
distance $R_{ij}$. The total number of holes is the same as the number 
of boron atoms; we have a {\bf half filled} band of interacting holes. 

On the insulating side of the above Hamiltonian, the low energy sector
is the spin sector, which can be seen clearly in the limit $U >> t$
as a random antiferromagnetic Heisenberg Hamiltonian with superexchange 
interaction $J_{ij}$:
\be
H_s  \approx  \sum_{ij} J_{ij} 
({\bf S}_i \cdot {\bf S}_j - \frac{1}{4}) 
\ee
This is well established in the context of Si:P for example leading to 
notions of hierarchy of singlet spin coupling, valence bond 
localization, valence bond glass etc., both experimentally and 
theoretically\cite{bhatLee}. We do not think diamond:B is fundamentally 
different from Si:P in the Mott insulating side.

The issue for us is the conducting side close to Mott transition point.
In the present paper, based on our recent study, we offer a new insight.
Conventionally the conducting side is thought of as a disordered fermi
liquid and with regions of certain local moment 
character\cite{bhatSachdev}. We have recently suggested\cite{gbrvbms} 
that a corresponding conducting state close to the Mott transition point 
in the non-random case is well thought of as a `self doped Mott insulator', 
a projected metal. In a self doped Mott insulator a small and 
{\bf equal density} of {\bf free} positive ($B^+$) and negative ($B^-$) 
carriers are self consistently generated across the Mott transition,
out of the neutral states $B^0$, in the process of minimizing 
the free energy, particularly the long range coulomb interaction part. 
Further, antiferromagnetic superexchange, that is characteristic of a 
Mott insulator survives in the conducting state as well, the same way 
superexchange survives in the $CuO_2$ planes of cuprates in the presence 
of doped holes. {\em That is, in the above conducting state, in addition 
to `virtual' double occupancy and empty sites, which are responsible 
for generation of superexchange, a small and equal density of `real' 
and delocalized double occupancy and empty sites are maintained}. 
Our work\cite{gbrvbms} unified RVB mechanism of superconductivity in 
hole doped cuprates with that in organic superconductors such as 2D ET 
salts and also predicted new systems.

An effective Hamiltonian of the above conducting state is the 2 species 
random t-J Hamiltonian that is obtained\cite{gbrvbms} by a superexchange 
perturbation theory for the present situation of half filled band of
disordered interacting electron Hamiltonian (equation 1)
\bearr
H_{tJ}  =  & - & \sum_{ij}t^{}_{ij}  P_d~c^\dagger_{i\sigma} 
c^{}_{j\sigma} P_d
     -  \sum_{ij}t^{}_{ij}  P_e~ c^\dagger_{i\sigma} c^{}_{j\sigma} P_e
 +  h.c.  \nonumber \\
      & - & \sum_{ij} J_{ij} ( {\bf S}_i \cdot {\bf S}_j - 
\frac{1}{4} n^{}_i n^{}_j ) + \sum_i \epsilon_i 
c^{\dagger}_{\sigma}c^{}_{\sigma}~, 
\eearr
operating in a subspace that contains a fixed number $xN$ of 
doubly occupied $B^-$ and $xN$ empty $B^+$ states. Here $N$ is the 
total number of electrons, which is the same as the number of lattice sites. 
The projection operators $P_d$ and $P_e$ allow for the hopping of  
charged $B^-$ and $B^+$ states respectively, maintaining $(1-2x)N$ singly 
occupied neutral states $B^0$ in a dynamical fashion.  Notice 
that the long range coulomb interaction has disappeared in the above 
Hamiltonian because of screening in the metallic state; More importantly 
it has left its effect parametrically through the generation of a finite 
density $2x$ of free carriers by the process of self doping.


Our 2 species random t-J model adapted to the self doped Mott insulator 
has a more
transparent form in the slave boson representation $ c^{\dagger}_{i\sigma}
\equiv s^{\dagger}_{i\sigma}d^{}_{i} + \sigma s^{}_{i{\bar \sigma}}
e^{\dagger}_{i} $. Here the charge-ons
$d^{\dagger}_i$'s and $e^{\dagger}_i$'s are hard core bosons that create
doubly occupied sites ($B^-$ of charge $e^-$)  and empty sites 
($B^+$ of charge $e^+$) respectively. The fermionic
spinon operators $s^{\dagger}_{i\sigma} $'s create singly occupied sites
($B^0$, charge neutral) with a spin projection $\sigma$. The local
constraint, $ d^{\dagger}_{i} d^{}_{i} + e^{\dagger}_{i} e^{}_{i} +
\sum_{\sigma} s^{\dagger}_{i\sigma} s^{}_{i\sigma} = 1 $, keeps us in 
the right Hilbert space.

In the slave boson representation our t-J model takes a suggestive form:
\bearr
 H_{tJ}  & = &  - \sum_{ij}t^{}_{ij}
( e^{\dagger}_{i} e^{}_{j} 
\sum_\sigma s^{\dagger}_{i\sigma} s^{}_{j\sigma} 
- d^{\dagger}_{i} d^{}_{j} \sum_\sigma 
s^{\dagger}_{i\sigma} s^{}_{j\sigma} ) + h.c.
\nonumber \\
& - & \sum_{ij} J_{ij} b^{\dagger}_{ij}b^{~}_{ij} + \sum_i \epsilon_i
(d^\dagger_{i} d^{}_{i} - e^\dagger_{i} e^{}_{i})  
\eearr
where $b^{\dagger}_{ij} = \frac{1}{\sqrt 2} ( s^\dagger_{i\uparrow} 
s^\dagger_{j\downarrow} -  s^\dagger_{i\downarrow} 
s^\dagger_{j\uparrow})$ is a spin singlet spinon pair creation 
operator at the bond $ij$. 

At a formal level, a slave boson analysis is readily performed for our 
random interacting electron model. We suggest two different phases,
a valence bond glass (VBG) and RVB superconducting phase, shown in 
figure 2.

\begin{figure}
\includegraphics*[width=8.0cm]{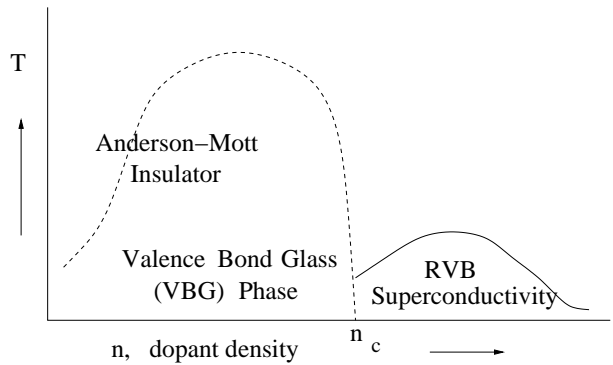}
\caption{\label{fig1} Schematic Phase Diagram as a function of
dopant density in Diamond:B, an uncompensated case.}
\end{figure}

In the Anderson-Mott insulator region we have a VBG phase. This is 
brought out well\cite{bhatLee} both in experiments and theory in 
the case of Si:P. This state contains only real `neutral' $B^0$ 
states and no {\bf real} $B^+$ and  $B^-$ charged states. 
A good variational wave function to describe such a valence bond 
glass state is
\be
|VBG\rangle =  \hat{P}~( \sum_{ij}\phi_{\rm VBG} (i,j)
b^{\dagger}_{ij} )^N |0\rangle , 
\ee
that describes condensation of neutral valence bonds in a specific
valence bond pattern given by the pair function $\phi_{\rm VBG}(i,j)$.
While one can formally write down a self consistent gap equation for
the pair function $\phi_{\rm VBG}(i,j)$ of our random correlated
spin-half system, one does not attempt to solve it but assumes that 
a solution exists.

In Si:P, the hierarchically organized valence bond couplings\cite{bhatLee}
 in a VBG  leads to a `pseudo gap' that gives a magnetic susceptibility that 
vanishes as $\chi_{\rm spin} \sim \frac{1}{T^{1-\alpha}}$ with $\alpha > 0$. 
The spin contribution to specific heat also gets a similar power law correction.

As mentioned earlier, across the Mott transition a small and equal
density of positive and negative carriers are spontaneously generated 
out of the insulating valence bond glass state. The delocalization 
of charged carriers causes charging and resonance of the frozen 
valence bonds resulting in an RVB superconducting state. This state 
is best described by a generalized RVB variational wave function, 
inferred from a slave boson mean field analysis:
\be
|RVB\rangle =  \hat{P}~ 
(e^{\dagger}_{\mu})^{xN} (d^{\dagger}_{\nu})^{xN}
(b^{\dagger}_{0})^{(1-2x)N} |0\rangle 
\ee
Here $e^{\dagger}_{\mu} \equiv \sum_i \phi^*_{\mu}(i) e^{\dagger}_i$
and  $d^{\dagger}_{\nu} \equiv \sum_i \phi^*_{\nu}(i) d^{\dagger}_i$
represent `bose condensation'\cite{krs} of the $B^+$ and $B^-$ in two different 
`extended states' $\phi_{\mu}(i)$ and $\phi_{\nu}(i)$. As the hopping 
matrix elements of $B^+$ and $B^-$ have opposite sign (equation 4), 
they condense in general in different extended states. The operator
$b_{0} \equiv \sum_{ij} \phi_{\rm SC} (i,j)b^{\dagger}_{ij}$ represents 
the condensation of valence bond pairs $B^0 -B^0 $ in an extended 
state represented by the pair function $\phi_{\rm SC}(i,j)$. And $\hat{P}$ 
is a projection operator that ensures the presence of only one of
$B^0$  or $B^+$ or $B^-$ acceptor states in any boron site in the 
many body wave function, by obeying the local constraint,
$ d^{\dagger}_{i} d^{}_{i} + e^{\dagger}_{i} e^{}_{i} +
\sum_{\sigma} s^{\dagger}_{i\sigma} s^{}_{i\sigma} = 1 $.   

To understand the above variational wave function (equation 6), we wish 
to state that in a non random case of simple cubic lattice for example,
the holons and doublons respectively condense at wave vectors (0,0,0) and 
$(\pi,\pi,\pi)$, in view of the different signs of the holon and doublon 
hopping matrix elements in equation 4. Also note that in the 
RVB theory, the holon and doublon condensation, an apparent charge `e' 
condensation\cite{krs} is actually a book keeping device\cite{pwaPRL87} 
for the charge `2e' condensation of physical electron pairs (valence bond 
pairs).

As mentioned earlier, in view of the random character of our Hamiltonian, 
one can only make some existence type of statement of the functions
$\phi_{\mu}(i)$,  $\phi_{\nu}(i)$ and $\phi_{\rm SC}(i,j)$ from 
plausibility arguments. For the same reason it is difficult to get
quantitative estimate of the superconducting \tc, from a formally
`exact' gap equation, even within the slave boson mean field analysis. 
However, we can get a very rough order of magnitude estimate of the 
superconducting \tc  using certain `typical' values of the impurity band 
parameters using a standard RVB mean field expression,  
$k_B$\tc $\approx \frac{W}{2} e^{-\frac{W}{J}}$. If we assume an impurity
band width $W$ around $0.2~eV$ and superexchange J around $0.04~eV$, 
we get a \tc in the range 1 to 10 K. It is important to remark that 
what is seen experimentally in diamond:B is likely to be some kind of 
lower bound for a larger intrinsic superconducting \tc, as inhomogeneities 
in the real system are likely to affect \tc considerably.

What is the order parameter symmetry ? Superconductivity is in a 
disordered lattice of boron acceptors.  Unlike cuprates, where 
a $d_{x^2-y^2}$-wave symmetry is compatible with an underlying square 
lattice symmetry, higher angular momentum states have no compatibility 
with the underlying random lattice. Hence, diamond:P is likely to go into 
an {\bf extended s-wave} superconducting state, there by also respecting 
Anderson's theorem on dirty superconductors.

The conducting side of the Anderson-Mott transition, in view of residual 
unscreened short range coulomb repulsion could create a charge density 
glass of the spontaneously created $B^+$ and $B^-$ carriers as well. 
This will be a competitor to superconductivity. 

If our correlation mechanism is at work, the normal state should be
an anomalous one. Disorder should lead to certain soft and localized 
spinon and chargeon quasi particle states. This is likely to lead to 
normal state anomalies different from cuprates.

Our current proposals, including mechanism of superconductivity, as it 
stands is very heuristic and based on limited available experimental 
results. To make further progress and check the validity and  
consequences of the present proposal, it is important to perform 
certain experiments: i) precise measurement of the critical 
concentration $n_c$ of the Anderson-Mott insulator to metal transition 
ii) map out the superconductivity phase diagram by changing the boron 
concentration iii) look for anomalous $T^\alpha$ type power law 
signals in specific heat and magnetic susceptibility in the insulating 
phase, characteristic of the valence bond glass phase and hierarchical
singlet coupling, iv) `spin gap' type of behaviour in the normal 
state on the conducting side, looking for onset of spinon pairing  
v) normal state transport anomalies on the conducting side and vi) look 
for superconductivity in the Mott insulator phase by small amount of 
compensation, say by nitrogen type donor impurities; this will be 
an external doping of the boron impurity band rather than self-doping.

Our present proposals raise some questions and suggests applicability
to other systems. Why superconductivity is absent in Si:P and related
systems ? Apart from the fact that the energy scales are low, the
impurity state orbital degeneracy is high and spin-orbit coupling is 
relatively high. Hund coupling and reduced band filling resulting from 
multiple bands is likely to diminish possibility of 
singlet superconductivity. It will be also interesting to look for 
new systems and also  find out
if electron correlation play any important role in estabilishing 
superconductivity in known impurity band superconductors\cite{hulm} 

I thank Prof S Maekawa for bringing to my attention reference [1], 
Prof. S Maekawa and Prof H Fukuyama for discussion, hospitality and 
IFCAM Visiting Professorship at Sendai. An informative discussion 
on spectrosopy of boron acceptor states in diamond with Prof A K Ramdas 
in February 2004 is acknowledged.

\end{document}